\documentclass[12pt]{article}
\usepackage{theorem,amsmath,amssymb,amsfonts} %

 \newcommand\eqlabel{\label} 
\newcommand{\pxd}[2]{S^{#1}_{#2}} %
\newcommand{\wienerp}[2]{W^{#1}_{#2}} %
\newcommand{\mprmin}[2]{\gamma^{(m)#1}_{#2}} %
\newcommand{\Qmin}[0]{\Q^{(m)}} %

\newtheorem{theorem}{Theorem}[section] %
\newtheorem{assumption}{Assumption}

\newtheorem{corollary}[theorem]{Corollary}

\newtheorem{definition}[theorem]{Definition}

\newtheorem{proposition}[theorem]{Proposition}

\theorembodyfont{\rmfamily}

\newtheorem{example}[theorem]{Example}
\newtheorem{remark}[theorem]{Remark}

\numberwithin{equation}{section}
\numberwithin{theorem}{section}
\newenvironment{proof}[1]{\medskip\noindent \textbf{Proof#1.~}}{\hbox{ }\hfill\vrule height6pt width6pt depth0pt \medskip} %

\renewcommand\i{\infty} %
\newcommand\R{\mathbb{R}}

\newcommand\E{\mathbf{E}}

\newcommand\IP{\mathbf{P}}
\newcommand\Q{\IQ}
\newcommand\IQ{\mathbf{Q}}

\newcommand\cF{\mathcal{F}}
\newcommand\cG{\mathcal{G}}

\newcommand\cM{\mathcal{M}}

\newcommand\p{\kern-1pt\cdot\kern-1pt} %

\newcommand\al{\alpha}

\newcommand\ga{\gamma}

\newcommand\ep{\varepsilon} %

\renewcommand\phi{\varphi}

\newcommand\arctg{\mathrm{arctg}}

\newcommand\subneq{\subseteq\!\!\!\!\!\raise-3pt\hbox{\tiny{$|$}}\,\,\,}%
\newcommand\supneq{\supseteq\!\!\!\!\!\raise-3pt\hbox{\tiny{$|$}}\,\,\,}%

\newcommand\JbarsigmaXY{{\overline J}^{\raise4pt\hbox{$\scriptscriptstyle\sigma(X,Y)$}}}

\newcommand\textfrac[2]{{\textstyle \frac{#1}{#2}}}

\date{October 10, 2007}

\title{In which Financial Markets do \\ Mutual Fund Theorems hold true?}

\author{W.~Schachermayer\thanks{wschach@fam.tuwien.ac.at;Vienna University of Technology, Wiedner Hauptstrasse 8-10/105, A-1040 Wien, Austria. Financial support from the Austrian Science Fund (FWF) under the grant P19456, from Vienna Science and Technology Fund (WWTF) under Grant MA13 and by the Christian Doppler Research Association (CDG) is gratefully acknowledged.} \and M. Sirbu  \thanks
{sirbu@math.utexas.edu; University of Texas at Austin,
  Department of Mathematics, 
1 University Station C1200,
Austin, TX, 78712. The research was
    supported in part by the National
    Science Foundation under Grant DMS-0604643. 
} \and E. Taflin\thanks{ taflin@eisti.fr; Chair in Mathematical Finance, EISTI,
Ecole International des Sciences du Traitement de l'Information,
Avenue du Parc, 95011 Cergy, France.}}

\begin{document}

\maketitle

\begin{abstract}
The Mutual Fund Theorem (MFT) is considered in a general semimartingale 
 financial market $S$
 with a finite time horizon $T$, where agents maximize expected utility 
of terminal wealth.
 It is established that: 
 
 \begin{enumerate}
\item  Let $N$ be the wealth process of the num\'eraire portfolio
(i.e. the optimal portfolio for the log utility).
If any path-independent option with maturity $T$ written on the num\'eraire portfolio can be replicated by trading \emph{only} in $N$, then 
the (MFT) holds true for general utility functions, and the num\'eraire 
portfolio  may serve as mutual fund. This generalizes Merton's classical result on Black-Scholes markets.

Conversely, under a supplementary weak completeness assumption, we show that the validity of the (MFT) for 
general utility functions implies the same replicability property for options on the num\'eraire portfolio described above.
 
\item  If for a given class of utility functions (i.e. investors) 
the (MFT) holds true in all complete Brownian financial markets $S$, then all 
investors use the same utility function $U,$ which must be of HARA type.
This is a  result in the spirit of the classical work by Cass and Stiglitz.
\end{enumerate}
\end{abstract}

\section{Introduction}

The \textit{Mutual Fund Theorem} (MFT), also sometimes called the ``two funds theorem'' or the ``separation theorem'', is one of the pearls of Mathematical Finance. Under suitable assumptions (for which we attempt to give a better understanding in the present paper), the optimal investment strategy of a utility maximizing agent has the following simple form:

\textit{The agent will only invest in two funds: the risk-free asset as well as  a second mutual fund  which is a linear combination of the risky assets available on the financial market. The crucial feature is that the same (second) mutual fund, i.e., the same linear combination (portfolio)  of risky assets applies to all utility maximizing agents, independently of the special form of their utility function as well as their initial endowment}.

This theme goes back to the work of Tobin \cite{Tobin 58}, who considered a cash and console bond market described by a mean-variance single-period model as analyzed in the work of Markowitz (\cite{Markowitz52}, \cite{Markowitz59}). Their original work focused on the analysis of the mean and variance of the asset returns and was restricted to the rather limited framework of single-period markets and quadratic utility functions.

Starting from these seminal papers, there were essentially two directions into which the subsequent research can be divided. On one hand,  Cass and Stiglitz \cite{Cass-Stiglitz 70} (see also \cite{Hakansson 69} and \cite{Ross78}) obtained negative results: for general discrete-time financial markets, the mutual fund theorem only holds true for very restrictive classes of utility functions (see Theorem \ref{propX2}  below for a result in this spirit). On the other hand, an important positive result was obtained by Merton (see \cite{Mert71} and \cite{Mert73}): in the framework of a continuous-time multi-dimensional Black-Merton-Scholes model, with deterministic drift and volatility coefficients, the mutual fund theorem does hold true (see Theorem 2 of \cite{Mert71} and Theorem 1 of \cite{Mert73}). The results of Merton were further developed: we refer here the reader to  \cite{Khanna-Kulldorff 99} and Chapter 3 Remark 8.9 of \cite{K-S 99} for generalizations 
 within the class of deterministic drift and  models, and to \cite{I.E.-E.T bond th} (see Theorem 3.10) for mutual fund theorems in the framework of  
 bond markets. 

In the present paper we want to develop a better understanding of this discrepancy and to obtain necessary and sufficient conditions (on  a financial model) for the mutual fund theorem to hold true. 
 Roughly speaking, MFT is directly related to the replicability of options written on the  num\'eraire portfolio process, so it is a completeness condition.  We recall that the num\'eraire portfolio is the optimal portfolio of a logarithmic utility maximizer, and we refer the reader to \cite{Becherer 2001} and \cite{K-K 06}  (and the references given there) for more details. If every (say, bounded)  European path-independent option on the num\'eraire portfolio which expires at the final time horizon $T$ can be replicated \emph{by trading only in the num\'eraire portfolio}, then the mutual fund theorem holds true with respect to arbitrary utility functions (see Theorem \ref{main1}) below. We present an example which shows that a direct converse of this theorem does not hold true (see Example 
 \ref{example2} below). Under proper conditions (which amount to a weak form of market completeness) we can also formulate a reverse result (Theorem \ref{main2} below). 

We emphasize that our findings not only apply to continuous price processes, but also to processes with jumps.
 We thus also get a better insight why in the setting of Cass and Stiglitz \cite{Cass-Stiglitz 70} things go wrong: there are only very few processes with   jumps generating a complete market. Essentially, the only cases are  the binomial model in discrete time and  
  the (compensated) Poisson process in continuous time. If the num\'erarie portfolio process has jumps and fails to be of this very special form (as is the case in \cite{Cass-Stiglitz 70}), then it follows as a very special case of Theorem  \ref{main2} that the mutual fund theorem does not hold true for arbitrary utility functions.

Our theorems apply to other cases of interest, e.g., some continuous models with stochastic volatility. The basic message is that the validity of the mutual fund theorem is an \emph{informational} issue, pertaining to the fact which kind of derivatives (in terms of their \emph{measurability} properties) can be replicated by only trading in a single mutual fund, which usually is the \emph{num\'eraire portfolio}.

Turning to the negative side, we can also provide some new insight . Proposition \ref{propX2}
shows that  the same negative assertions as in Cass and Stiglitz \cite{Cass-Stiglitz 70}, in the framework of discrete-time models, can be proved in the framework of continuous processes in continuous time. 

We would like to thank Dmitry Kramkov and Steven Shreve for their interesting remarks made during the ICMS workshop, Edinburgh, July 9-13, 2007 and the Mathematical Finance Seminar at Carnegie Mellon.

\section{The Mathematical Model }

We consider a financial market, on a finite time interval $[0,T]$, with one riskless asset $S^0$ called the bond (or better, money market account) and   $d$ risky assets called stocks. We choose $S^0$ as  num\'eraire (which means we normalize $S^0=1$) and  denote  by $\pxd{1}{}, \ldots, \pxd{d}{}$ the prices of the risky assets measured in units of $S^0$.
The price process of the
stocks $S=(S^i)_{1\leq i\leq d}$ is assumed to be a \emph{locally bounded} semimartingale on
a filtered probability space $(\Omega, \mathcal{F} , (\mathcal{F}
_t)_{0\leq t\leq T}, \mathbb{P})$, where  the filtration $(\cF _t)_{0\leq t\leq T}$ satisfies the \emph{usual conditions} (right continuous and saturated)
and $\mathcal{F} = \mathcal{F}_T$.

A portfolio is defined as a pair $(x,H)$, where the constant $x$
represents the initial capital and $H=(H^i)_{1\leq i\leq d}$ is a $(\cF_t)_{0\leq t \leq T}$-predictable 
and $S$-integrable process in the vector integration sense.  The wealth process
$X=(X_t)_{0\leq t\leq T}$ of the portfolio evolves in time as the
stochastic integral of $H$ with respect to $S$:
\begin{equation}
  \eqlabel{eq:portfolio}
  X_t = x + \int_{0} ^{t}( H_u ,dS_u),\quad 0\leq t\leq T.
\end{equation}
For each $x>0$ we denote by $\mathcal{X}(x)$ the family of wealth processes $X=(X_t)_{0\leq t\leq T}$ with
nonnegative capital at any instant and with initial value equal to
$x$:
\begin{equation}
  \eqlabel{eq:X(x)}
  \mathcal{X}(x) \triangleq \{X\geq 0: ~  X \mbox{ is defined by }
  (\ref{eq:portfolio})\}.
\end{equation}
A probability measure $\Q\sim \IP$ is called an
\emph{equivalent local martingale measure} if  the locally bounded semimartingale $S$ 
is a local martingale under $\Q$.  We denote by $\mathcal{M}^e$
the set of equivalent local martingale measures and assume, as usually, that
\begin{assumption}\eqlabel{ass1}
\begin{equation}\eqlabel{eq:NA}
  \mathcal{M}^e \neq \emptyset.
\end{equation}
\end{assumption}
We would like to point out that all the results in the present paper can be proven without the local boundedness assumption on $S$, by changing slightly the definition of equivalent martingale measures (the technical details related to the case of unbounded semimartingales are  described and analyzed in \cite{DS}). We decided to avoid here unnecessary technicalities by assuming that the stock price process $S$ is \emph{locally bounded}.
 
In this financial model we consider economic agents whose
preferences are modeled by \emph{expected (deterministic) utility from terminal wealth}. The cases of \emph{expected utility from consumption} as well as \emph{random utility functions} are left to forthcoming work.
A  generic  (deterministic) utility function 
$U:(0,\infty)\rightarrow \R$ is
assumed to be strictly concave, continuously differentiable and strictly increasing. In addition, it satisfies  the Inada conditions
\begin{equation} \eqlabel{Inada}
\lim_{x \searrow 0} U^{\prime }( x) \rightarrow \infty \quad \text{and} \quad \lim_{x\rightarrow \infty} U^{\prime }( x) =0,
\end{equation}
as well as the \textit{reasonable asymptotic elasticity condition} (see \cite{Kr-Scha 99}),
\begin{equation} \eqlabel{InadaF}
\limsup_{x \rightarrow \infty} xU'(x)/U(x)<1.
\end{equation}

We resume this in the subsequent statement.

\begin{assumption} \eqlabel{ass2}
The utility function $U:(0,\infty) \rightarrow \R $ is strictly increasing, strictly concave and differentiable on $(0, \infty)$ and formulas \eqref{Inada} and \eqref{InadaF} hold true.
\end{assumption}
We now  introduce the problem of \emph{optimal investment from  terminal wealth} for an economic agent and define the \emph{indirect utility function} $u$ by:
\begin{equation}\eqlabel{sup1}
u(x)=\sup_{X \in \mathcal{X}(x)} \E[U(X_T)], \ x>0.
\end{equation}
\begin{assumption}\eqlabel{ass2'}
We assume that $u(x)<\infty$, for some $x>0$ (or, equivalently, for all $x>0$).
\end{assumption}
\begin{remark}\eqlabel{rem:more-utilities} The problem of optimal investment from terminal wealth \eqref{sup1} can be studied for different classes of utility functions.
For example, one can consider utility functions 
$U:(a,\infty)\rightarrow \mathbb{R}$, for some finite $a$. 
In this case, the Inada conditions have the form 
\begin{equation} \eqlabel{Inada'}
\lim_{x \searrow a} U^{\prime }( x) \rightarrow \infty \quad \text{and} \quad \lim_{x\rightarrow \infty} U^{\prime }( x) =0,
\end{equation}
and the optimization problem can be regarded as a translation with respect to $x$ of an optimization problem for $a=0$. All the  results related to the optimization problem \eqref{sup1}  and our results below on Mutual Fund Theorems hold true modulo some obvious modifications: for example, in Proposition \ref{propX2}
the utilities would be translations in $x$ of affine transformations of power utilities, etc.

On the other hand, the case of utilities which  are finite-valued on the whole real line, 
$U:\R \rightarrow \mathbb{R}$, such as the exponential utility $U(x)=-e^{-x}$,  requires a slightly different duality approach. 
For the sake of presentation, we decided to not include this case here: however, we want to point out that our results carry over to this setting too.

\end{remark}
We denote by $\hat{X}(x,U)$ the optimal wealth process in \eqref{sup1}, which exists and is unique under  Assumptions \ref{ass1}, \ref{ass2} and \ref{ass2'} (see \cite{Kr-Scha 99} ). 

A particularly important  utility function $U$ is the \textit{logarithmic utility}
\begin{equation} \eqlabel{log U}
U(x)=\ln (x), \quad x > 0.
\end{equation}
In this case, Assumption \ref{ass2'} takes the form 
\begin{assumption}\eqlabel{ass-log}
Assume that 
\begin{equation}\eqlabel{eq:finite-log}
\sup_{X \in \mathcal{X}(1)} \E[\ln(X_T)]<\infty.
\end{equation}
\end{assumption}
Assumption \ref{ass-log}, which is an assumption on the model,  is equivalent to the dual condition
\begin{equation}\eqlabel{finite-dual-log}
\inf_{\Q \in \mathcal{M}^e} \E[-\ln(\frac{d\Q}{d\IP})]<\infty.\end{equation}
The equivalence between Assumption \ref{ass-log} and \eqref{finite-dual-log} follows from  the duality arguments in \cite{Kr-Scha 99}. Some of these duality arguments  are briefly described in Section \ref{proofs} below, since they are needed to proof our own results.

Assuming that Assumption \ref{ass-log} is satisfied, we denote by $N$ the optimal wealth process for the logarithmic maximizer with initial endowment $1$, i.e.\ $N=\hat{X}(1, \ln)$. The wealth process $N$ is then the well-known \textit{num\'eraire portfolio} process (we refer the reader to \cite{Becherer 2001} and \cite{K-K 06} for a thorough analysis). The num\'eraire portfolio (which can even be defined in the absence of either the local boundedness assumption on $S$ or Assumption \ref{ass-log} ) can be caracterized as the unique num\'eraire that makes any positive wealth process (measured in terms of  this num\'eraire) a supermatingale  under the original probability measure $\IP$. 
In particular, the constant wealth process $X_t=1$ for $0\leq t \leq T$, measured in terms of the num\'eraire $N$, is a supermatingale, which means that the process $Z$ defined by
$$Z_t=\frac{1}{N_t},\ \ 0\leq t\leq T,$$
is a supermartingale under $\IP$.
\begin{assumption}\eqlabel{ass1'}
The process $Z$ is a martingale, i.e
$$\E[Z_T]=1.$$
\end{assumption}
If Assumption \ref{ass1'} is satisfied,  we can define the probability measure $\Qmin \sim \IP$ by
$$\frac{d\Qmin}{d\IP}=Z_T.$$ 
Under our standing assumption that $S$ is locally bounded, the measure $\Qmin$ is an \emph{equivalent martingale measure}, i.e $\Qmin \in \mathcal{M}^e$. In particular, each \emph{bounded} wealth process is a \emph{martingale} under $\Qmin$.
For the case of continuous stock price process $S$, the measure $\Qmin$  was introduced in \cite{Foll-Schw1991}, Definition 3.2 under the name of \emph{minimal martingale measure}.
\begin{remark}\eqlabel{rem:Brownian} Our semimartingale market model includes the particular case of
models driven by Brownian Motions.  We describe this  briefly:  suppose that the price process $(\pxd{}{t})_{ 0\leq t\leq T}=(\pxd{1}{t}, \ldots, \pxd{d}{t})_{ 0\leq t\leq T}$ satisfies the stochastic differential equation (SDE)
\begin{equation}\eqlabel{eqS}
 \frac{d\pxd{i}{t}}{\pxd{i}{t}}= \sum_{1 \leq j \leq N} \sigma _t ^{i,j}(\gamma^{j}_t \, dt +d\wienerp{j}{t}), ~~~ 1 \leq i \leq d
\end{equation} where $\wienerp{}{}=(\wienerp{1}{}, \ldots, \wienerp{N}{})$ is a standard $N$-dimensional Brownian motion with respect to the filtered probability space $(\Omega, \IP, \cF_T, (\cF_t)_{ 0\leq t\leq T})$, where $(\cF_t)_{ 0\leq t\leq T}$ is a right continuous and saturated filtration.
The $d \times N$-matrix-valued volatility process $\sigma$ with matrix elements $\sigma^{ij}_{t}$ at time $0\leq t\leq T$ and the $N$-dimensional market price of risk process $\gamma =
(\gamma _t)_{0\leq t\leq T}$, where $ \gamma _t=(\gamma^{1}_t, \ldots, \gamma^{N}_t)$, are assumed to be optional with respect to $(\cF_t)_{ 0\leq t\leq T}$ and locally square  integrable.
In general, the market price of risk $\gamma_t$ is not uniquely determined by the process $S$ via \eqref{eqS}. 
Indeed, two processes  $(\ga_t)_{0\leq t\leq T}$, and 
 $(\tilde{\ga}_t)_{0\leq t\leq T}$  such that 
 $\ga_t-\tilde{\ga}_t\in\ker(\sigma_t)$ a.s, for all $0\leq t\leq T$ 
 define the same process $S$ via \eqref{eqS}.
 There is a unique  $\ga$ such  that $\ga_t$ is a.s.\ orthogonal to the kernel of $\sigma_t$, for all $0\leq t\leq T$. 
We denote it by $\mprmin{}{}$ and call it  \textit{Minimal Market Price of Risk} (MMPR).
The  num\'eraire portfolio process $N$ and its inverse process $Z$ can be computed in this case in terms of the minimal market price of risk:
\begin{equation} \eqlabel{SDE opt X log}
dN_t=N_t\left( (\mprmin{}{t}, d\wienerp{}{t}) + \textfrac{1}{2}\|\mprmin{}{t}\|^{2}dt\right), ~0\leq t\leq T, ~N_0=1,
\end{equation}
and 
\begin{equation}\eqlabel{xi-min}
Z_t=\exp\left( -\int_{0}^{t}(\mprmin{}{s}, d\wienerp{}{s}) -\frac{1}{2}\int_{0}^{t} \|\mprmin{}{s}\|^{2}ds\right), ~~~ 0\leq t\leq T.
\end{equation}
 \end{remark} 

Going back to  general semimartingale models, we now can formalize the idea of the Mutual Fund Theorem. In order to do this we need two definitions:
\begin{definition}\label{MF} We call a mutual fund any wealth process $M$ with initial capital equal to one, i.e any 
$$M_t=1+\int _0^t (H^M_u, dS_u),\ \ 0\leq t\leq T,$$
where $H^M=(H^M_t)_{0\leq t\leq T}$ is an $\R ^d$-valued predictable and $S$-integrable process.
\end{definition}
\begin{definition} \label{def MFT}
Let the financial market $\pxd{}{}$ satisfy Assumption \ref{ass1}  and let $\mathcal{U}$ be a family of utility functions satisfying Assumptions \ref{ass2}, \ref{ass2'}. We say that the financial market $\pxd{}{}$ satisfies the the Mutual Fund Theorem (MFT) with respect to $\mathcal{U}$  if there exists
a mutual fund $M$ 
such that
\begin{equation}
\forall \; U \in \mathcal{U},  x>0  \text~ \exists \; k=k(x,U)
\end{equation}
such that
\begin{equation} \eqlabel{eq MFT}
\hat{X}_t(x,U)=x+\int _0^t k_u dM_u,\ 0\leq t\leq T,
\end{equation}
where $k=k(x,U)$ is a real-valued $(\cF_{t})_{0\leq t\leq T}$ predictable, $M$-integrable process.
The process $M$  is then called a mutual fund for the market $\pxd{}{}$ and the class of utility functions~$\mathcal{U}$.
\end{definition}

The interpretation  of \eqref{eq MFT} is that the optimal investment strategy is to invest in the mutual fund $M$ and  the money market only.
In the case when the logarithmic utility function $U(x)=\ln x$ is in $\mathcal{U}$  
we will show that one may choose the \emph{num\'eraire portfolio}  as mutual fund, i.e.
$M=N$, under a mild  technical assumption (according to Remark \ref{rem:mf-needed} below).
\section{Results}\eqlabel{results}
In what follows, for any stochastic process $X=(X_t)_{0\leq t\leq T}$,  we will denote by $(\mathcal{F}^X_t)_{0\leq t\leq T}$ the right-continuous and saturated filtration generated by the  process $X$, and by $\cG^X_t=\sigma (X_t)$ the sigma-algebra generated by the random variable $X_t$ 
 for a fixed $0\leq t \leq T$. Therefore, the space  $L^{\infty}(\Omega, \cG^N_T, \IP)$ or, briefly, $L^{\infty}(\cG^N_T)$ 
 can be interpreted as the collection of all bounded path-independent options on the num\'eraire portfolio $N$ expiring at time $T$:
$$L^{\infty}(\cG^N_T)=\{h(N_T),\textrm{~where~} h:\R\rightarrow \R \textrm{~is a bounded Borel measurable function}\}.$$
All along the paper, we will simply call \emph{European option} any path-independent option with fixed maturity (e.g. random variables in $L^{\infty}(\cG^N_T)$ are European options on $N$ with maturity $T$). We now define $\mathcal{R}(S)$ to be  the set of all
bounded random variables that are replicable by trading in the market $S$, i.e. the set of  bounded random variables $f$  for which there exists a number $p(f)$ and $d$-dimensional   $(\mathcal{F}_t)_{0\leq t\leq T}$ -predictable
and $S$-integrable process $H^f$ such that 
$$f=p(f)+\int _0 ^T ( H^f_u,dS_u),$$
and the stochastic integral is a bounded process.  The process $P(f)=(P(f))_{0\leq t\leq T} $ defined by 
$$P_t(f)=p(f)+\int _0 ^t ( H^f_u,dS_u),\ \ \ 0\leq t\leq T$$
is called the price process of $f$, and $p(f)$ the initial price of $f$. For a fixed mutual fund $M$ (see Definition \ref{MF}) we also define 
$\mathcal{R}(M) $ to be the set of bounded random variables which can be replicated by trading in this mutual fund  only, i.e.
\begin{equation}\eqlabel{eq:replM}
f=p(f)+\int _0^T k^f_udM_u,
\end{equation}
where $k^f$ is a one-dimensional  $(\mathcal{F}_t)_{0\leq t\leq T}$ -predictable process, which is $M$-integrable, and such that the stochastic integral is bounded.   The  set  of random variables $f\in \mathcal{R}(M)$ where the integrand $k^f$ in 
\eqref{eq:replM} can be chosen to be $(\mathcal{F}^M_t)_{0\leq t\leq T}$-predictable is denoted by $\mathcal{R}(M,\cF^M)$. It is obvious that
$$\mathcal{R}(M,\mathcal{F}^M)\subset \mathcal{R}(M)\subset \mathcal{R}(S),$$
no matter what mutual fund $M$ we choose.
Finally, for a fixed model $S$ satisfying Assumption \ref{ass1}, we denote  by $\mathcal{A}(S)$ the set of all utility functions satisfying 
Assumptions \ref{ass2} and \ref{ass2'}. We emphasize that the class $\mathcal{A}(S)$  depends on the model $S$, due to Assumption \ref{ass2'}.
We can now state our first main result, which provides \emph{sufficient conditions} for the validity of the mutual fund theorem:
\begin{theorem} \eqlabel{main1} 
Let the semimartingale financial market $S$ satisfy Assumptions \ref{ass1} , \ref{ass-log} and \ref{ass1'}.
If there is a mutual fund $M$ such that  each  bounded  European option $f$  with maturity $T$  and written on the num\'eraire portfolio $N$ can be replicated by trading only in the mutual fund $M$, then the financial model $S$ satisfies the Mutual Fund Theorem for the class  $\mathcal{A}(S)$ of all utility functions satisfying Assumptions \ref{ass2} and \ref{ass2'} (and $M$ is the corresponding mutual fund).

Speaking more formally, if the one-dimensional replicability condition 
$$(\bf{R})\ \ \ \ \ \ \ \ \ \ \ \ \ \  \ \ \ \textrm{there exists a mutual fund}\ M\ \textrm{such that~} L^{\infty}(\cG^N_T) \subset \mathcal{R}(M),\ \ \ \ \ \ \ \ \ \ \ \ $$ is satisfied  then 
$\pxd{}{}$ satisfies the (MFT) with respect to  the class $\mathcal{A}(S)$ of utility functions.
\end{theorem}
Theorem \ref{main1} has the following obvious consequence:
\begin{corollary} \label{cor:completeness}
Let the financial market $S$ satisfy Assumptions \ref{ass1}, \ref{ass-log} and \ref{ass1'}.
If the num\'eraire portfolio process $N$ defines a complete market with respect to its own filtration
$(\mathcal{F}^N_t)_{0\leq t\leq T}$, then the financial model $S$ satisfies the (MFT)  with respect to the set of all utility functions $\mathcal{A}(S)$, and the num\'eraire portfolio $N$  can be chosen as  Mutual Fund.
\end{corollary}
There is no direct converse to  Theorem \ref{main1}, i.e. the replicability condition ({\bf R}) is not necessary for the mutual fund theorem to hold true. Indeed, if the financial model $S$ is one-dimensional then the mutual fund theorem trivially holds true (the only stock may serve as mutual fund)
while the  replicability condition ({\bf R}) may fail, as explained in Example \ref{example2}.
Therefore, in order to get a get an implication in the opposite direction, we need to impose some extra assumption on the model. The  assumption is  actually a weak form of completeness, requiring that
any  European option on the num\'eraire portfolio $N$ can be hedged by trading in all stocks, and it can be expresses as:
$$({\bf WC})\ \ \ \ \  \ \ \ \ \ \ \ \ \ \ \ \ \ \ \ \ \ \ \ \  L^{\infty}(\cG^N_T)   \subset \mathcal{R}(S).$$
Assuming  ({\bf WC}), indeed, the mutual fund theorem and the replicability condition ({\bf R}) are equivalent, so we can (informally) summarize the main result(s) of the paper as
$$({\bf R}) \Longleftrightarrow ({\bf MFT})+({\bf WC}).$$
More explicitly, we formulate below the actual theorem on  \emph{necessary conditions} for the validity of the mutual fund theorem:
\begin{theorem}\eqlabel{main2} 
 Let the  financial market $S$ satisfy Assumptions \ref{ass1} , \ref{ass-log} and \ref{ass1'} as well as the weak completeness condition ({\bf WC}). 
 If $\pxd{}{}$ satisfies the (MFT) with respect to   the class $\mathcal{A}(S)$ of all utility functions satisfying Assumptions \ref{ass2} and \ref{ass2'}  then  the replicability condition ({\bf R}) holds true.
\end{theorem}
\begin{remark}\eqlabel{suff-cond-wc}
The hypothesis ({\bf WC}) of Theorem \ref{main2} is certainly satisfied if 
$S=(S_t)_{0\leq t\leq T}$ is a complete model with respect to the original filtration 
$(\mathcal{F}_t)_{0\leq t\leq T}$. Theorem \ref{main2} can also be applied in some incomplete markets using the idea of completion by "fictitious securities" described in \cite{KLSX}. For example, in the Brownian framework of Remark \ref{rem:Brownian}, if the filtration is  generated by the Brownian motion $W$, we can add a finite number of securities to create a complete market that admits as the unique martingale measure the measure $\Qmin$. The enlarged model is complete, so we can apply Theorem \ref{main2} to the new model. If no utility maximizer (for some $U\in \mathcal{A}(S)$) chooses to invest in the "fictitious securities", then Theorem \ref{main2} applies to the original incomplete model. As a matter of fact, the possibility of completing the market in such a peculiar way is characterized in Theorem \ref{max-stoch-dom} below and means that the dual optimizer is the same for all utility maximizing agents, as described in Section \ref{proofs}, Remark \ref{rem:stoch-dom}. 
\end{remark}
In this spirit we give   a dual characterization of the weak completeness condition ({\bf WC}).
We recall  that for two positive random variable $\xi$ and $\zeta$ we say that $\xi$ dominates $\zeta$ stochastically in the second order and write $\xi \succeq _2 \zeta$ if 
$$\int _0^t \IP[\xi \geq u]du \geq \int _0^t \IP[\zeta \geq u]du \textrm{~for each~}t\geq 0.$$
\begin{theorem}\eqlabel{max-stoch-dom} Assume that the financial model  $S$ satisfies Assumptions \ref{ass1}, \ref{ass-log} and \ref{ass1'}. Then the following statements are equivalent:
\begin{enumerate}
\item \textrm{the weak completeness condition ({\bf WC}) holds true}
\item for each $\Q\in \cM^e$ we have 
\begin{equation}\eqlabel{eq:trace}
\E[\frac{d\Q}{d\IP}|\cG^N_T]=\frac{d\Q^{(m)}}{d\IP}
\end{equation}

\item for each $\Q\in \cM^e$ we have
\begin{equation}\eqlabel{eq:stoch-dom}
\frac{d\Q^{(m)}} {d\IP} \succeq _2\frac{d\Q}{d\IP}.
\end{equation}
\end{enumerate}
\end{theorem}
 \begin{remark}\eqlabel{rem:ubp} Instead of the optimal investment problem \eqref{sup1} one  can  consider the more general problem of \emph{optimal investment with random endowment}:
 \begin{equation}\eqlabel{sup2}
u(x,q;f)=\sup_{X \in \mathcal{X}^b(x)} \E[U(X_T+qf)],
\end{equation}
where $f$ is a bounded random variable on $(\Omega, \IP, \mathcal{F})$ having the economic meaning of a contingent claim (option), $q\in \R$ represents the number of options and $\mathcal{X}^b(x)$ is the set of \emph{bounded} wealth processes starting at $x$. We refer the reader to \cite{Cvit-Sch-Wang 01} and \cite{Hug-Kram 04} for a detailed treatment of the duality theory related to this optimization problem.
For fixed $(x,q)$ such that $-\infty <u(x,q)<\infty$ one can define the \emph{utility-based price} $p(x,q;f)$ of the claim as any price $p$ such that
$$u(x+\tilde{q}p,q-\tilde{q};f)\leq u(x,q;f)\ \textrm{~for any~} \tilde{q}\in \R.$$
In other words, $p(x,q)$ is a price set in such a way that, if the investor having $x$ initial capital and $q$ contingent claims is allowed to trade at time zero the claims for the price $p(x,q;f)$, the optimal strategy is to keep the $q$ claims and trade only in stocks.
The price $p(x,0;f)$, analyzed extensively in the literature (for example  in  \cite{Davis}, \cite{F}, \cite{H}, \cite{Rubinstein}) and sometimes called Davis price, represents the partial equilibrium price of the claim corresponding to zero demand.  Using refined duality arguments, it is proved in \cite{HugonKramSch:04} that the price $p(x,0;f)$ is uniquely defined and it can be computed as the expectation of the claim $f$ under the \emph{dual minimizer measure} in case that such  minimizer actually exists.  
   In other words
   $$p(x,0;f)=\E _{\Q(x,U)}[f],$$
   where $\Q(x,U)$  (the \emph{pricing measure}  of the agent with utility $U$, initial capital $x$ and zero demand for the claims) is the minimizer in \eqref{alternative} below, in case there is a minimizer.    Taking the above discussion into account,   the completeness condition ({\bf WC}) is equivalent (under suitable technical conditions) to each of the following economically meaningful conditions
   \begin{itemize}
   \item all economic agents, independent of their utility function $U$ and initial capital $x$ have the same pricing measure:
   $$\Q(x,U)=\Qmin \textrm{~for all~} U\textrm{~and~}x$$
   \item all economic agents,  independent of their utility function $U$ and initial capital $x$
   assign the same prices to  infinitesimally small quantities of contingent claims i.e.
   $p(x,0;f)$ does not depend on $U$ and $x$, for each $f \in L^{\infty}(\Omega, \mathcal{F},\IP)$.
      \end{itemize}
In addition, it was shown in \cite{Kr-Si 06} that the \emph{second order stochastic domination} condition \eqref{eq:stoch-dom} is equivalent to the validity of some important qualitative properties of  first order approximations of utility based prices $$p(x,q;f)=p(x,0;f)+q\frac{\partial p}{\partial q}|_{q=0}+o(q),$$ so the completeness condition ({\bf WC}) relates directly not only to the behavior of utility-based prices in the zero order (i.e. the Davis price $p(x,0;f)$) but also to the first order, which captures most of the \emph{nonlinearity} of the pricing rule derived by the classical principle of \emph{marginal rate of substitution}.
  \end{remark}
\begin{remark}\eqlabel{rem:mf-needed}A natural question is whether  Theorems \ref{main1} and \ref{main2}  be reformulated \emph{only} in terms of the num\'eraire portfolio $N$, without any need of a (possibly different) mutual fund $M$. In other words, if we define the (\emph{stronger than} ({\bf R})) replicability condition 
$$(\bf{R^N})\ \ \ \ \ \ \ \ \ \ \ \ \ \  \ \ \ \ L^{\infty}(\cG^N_T) \subset \mathcal{R}(N),\ \ \ \ \ \ \ \ \ \ \ \ $$ 
do we have the equivalence 
$$({\bf R^N})\Longleftrightarrow({\bf MFT})+({\bf WC})?$$
The answer is no, in general. The rather simple Example \ref{ex:mf-needed} below shows that we can have a complete market, where the mutual fund theorem holds, but the num\'eraire portfolio cannot be chosen to be the mutual fund. In other words, while $({\bf R^N})\Longrightarrow({\bf MFT})+({\bf WC})$ always hods true (in an obvious way, taking into account Theorem \ref{main1}), the implication $({\bf MFT})+({\bf WC})\Longrightarrow ({\bf R^N})$ (corresponding to Theorem \ref{main2}) may fail. However, under a mild technical assumption, we can show that the  num\'eraire portfolio \emph{can be} chosen as mutual fund.
  In the Brownian framework of Remark
\ref{rem:Brownian}, the technical  hypothesis can be formulated as "the minimal market price of risk never vanishes", i.e.
\begin{equation}\eqlabel{eq:mpr-zero}
\gamma ^{(m)}_t\not= 0 \textrm{~whenever~} \sigma _t\not= 0\textrm{~for~} d\lambda \times \IP \textrm{~a.e.~} (t,\omega )\in [0,T]\times \Omega.
\end{equation}
In the general semimartingale framework, a similar condition, informally written as 
"$dN_t\not = 0$ whenever $dS_t\not =0$" should be imposed. The rigorous version of such a condition involves technical details on semimartingales (see, for example \cite{JS}) which  we choose not to formulate  here. 
\end{remark}
A second natural question, related to  \emph{information} and motivated by Corollary \ref{cor:completeness},  is the following:
 is the condition $L^{\infty}(\cG^N_T) \subset \mathcal{R}(N)$ equivalent to the stronger one
$$  L^{\infty}(\cG^N_T)    \subset \mathcal{R}(N, \mathcal{F}^N) ?$$ 
In other words, if any bounded European option on $N$ can be hedged  by trading in $N$ only, but using the full information $\mathcal{F}$, can we do this by just using the information $\mathcal{F}^N$?
 In terms of mutual funds, the question amounts to: if any rational investor will only invest in the mutual fund corresponding to the logarithmic maximizer, can she do so only observing the evolution of this mutual fund and having no other information about the market?
The answer is negative in general, and will be summarized in the following result:
\begin{proposition}\label{prp}  There exists a complete financial market $S$ 
generated by a two-dimensional Brownian Motion $W$ as in \eqref{eqS} and satisfying Assumptions \ref{ass1}, \ref{ass-log} and \ref{ass1'} such that
\begin{enumerate}
\item the original filtration $(\cF_t)_{0\leq t\leq T}$ is the filtration generated by the driving Brownian Motion
$(\cF^W_t)_{0\leq t\leq T}$, and, in addition, it equals the filtration generated by the stock price process $S$, i.e.
$$(\cF_t)_{0\leq t\leq T}=(\cF ^W_t)_{0\leq t \leq T}=(\cF  ^S _t)_{0\leq t\leq T} $$
\item any bounded path-dependent option on the num\'eraire portfolio $N$ can be replicated by only trading in $N$ and observing the "full" information $(\cF _t)_{0\leq t\leq T}$, i.e.
$$L^{\infty}(\mathcal{F}^N_T) \subseteq \mathcal{R}(N),\textrm{~hence, in particular~}L^{\infty}(\cG^N_T) \subseteq \mathcal{R}(N)$$
\item there exists a (path-independent) European option on the num\'eraire portfolio $N$ with maturity $T$  that cannot be replicated by trading in $N$ and using only the information obtained by observing the process $N$. In other words
$$L^{\infty}(\cG^N_T) \nsubseteq \mathcal{R}(N,\cF^N).$$
\end{enumerate}
\end{proposition}

Also on the negative side,
 we show that, if we allow for general Brownian market models as described in Remark \ref{rem:Brownian}, we cannot hope for a reasonable Mutual Fund Theorem to hold true. In fact, Proposition \ref{propX2} below gives an analogous result to  Cass and Stiglitz \cite{Cass-Stiglitz 70} obtained for processes in discrete time.
 
Let us first recall that the optimization problem \eqref{sup1} for either $U(x)=\ln x$ or $U(x)=\frac{x^\alpha}{\alpha}$, where $\alpha \in \; ]-\infty,1[ \; \setminus \{0\}$ has an obvious scaling property holds true, namely 
\begin{equation} \eqlabel{scaling}
\hat{X}(x,U)=x\hat{X}(1,U),\ x>0,
\end{equation}
as long as such a utility function  provided  satisfies Assumption
\ref{ass2'} . As a matter of fact, the only  $C^2$ utility function, for which the scaling property \eqref{scaling} holds true, are those obtained by the transformation 
\begin{equation}\label{aff_transf}
U\rightarrow A+B U,
\end{equation} where $A \in \R$, $B>0$ and $U$ is either the logarithmic or power utility.

In order to state the next result, which is related to the seminal papers \cite{Cass-Stiglitz 70} and \cite{Hakansson 69}, we need a technical strengthening of the regularity conditions on the utility function $U$:
\begin{assumption}\label{ass2f}
The utility function $U$ is $C^3$ on its domain $(0,\infty)$. 
\end{assumption}

\begin{theorem}\label{propX2}
Let $\mathcal{U}$ be a class of utility functions satisfying Assumptions \ref{ass2} and \ref{ass2f}.
Assume that  every 
Brownian 
financial market $S$  (as described in Remark \ref{rem:Brownian}) with values in $\R^2$, satisfying Assumption \ref{ass1} and also Assumption \ref{ass2'} with respect to all utilities $U\in \mathcal{U}$,
satisfies (MFT) with respect to the class of utilities 
$\mathcal{U}$. 

Then the family $\mathcal{U}$ consists only of a single utility function $U$ (modulo affine transformations of the type ( \ref{aff_transf}))  which is eihter
\begin{enumerate}
\item $U(x) = \log(x), \ x>0$ or
\item $U(x) = \frac{x^\al}{\al},\ x>0$, for some $\al\in ]-\i,1[\setminus \{0\}$.

\end{enumerate}
\end{theorem}

\begin{remark}
The above Theorem shows that a class of investors  would all invest in the same mutual fund, independent of their initial wealth and in \emph{any} financial model, if and only if they all have the \emph{same constant relative risk-aversion coefficient} (in the case when preferences of investors are described by utility functions which are finite for $x>0$).
If $\mathcal{U}=\{U\}$ (singleton), where the utility function $U$ is of logarithmic or power type, then by the scaling argument \eqref{scaling}
the mutual fund theorem trivially holds true for any model $S$. The message of Theorem  
\ref{propX2} is that this trivial  case is the only possible one, once we allow for general financial models of the form described in Theorem \ref{propX2}.

 A similar result to Theorem
\ref{propX2}  holds true if we consider utilities $U:(a,\infty)\rightarrow \R$,
or even $U:\R \rightarrow \R$. In the former case, the conclusion of Theorem \ref{propX2} is just a translation on the $x$-axis, and in the latter,  the class $\mathcal{U}$ is generated by affine transformations of 
$$U(x) = -\exp(-\al x),\ x\in \R,$$
\end{remark}
for some $\al>0$. The proofs require slightly different technical details, as pointed out in Remark \ref{rem:more-utilities}.
\section{Proofs and examples}\eqlabel{proofs}

A well known tool in studying the optimization problem
(\eqref{sup1})  is the use of duality relationships in the spaces of
convex functions and semimartingales.  Following \cite{Kr-Scha 99}, for a fixed utility function $U$, we  define the dual optimization problem to \eqref{sup1} as follows
\begin{equation}
  \eqlabel{eq:dual}
  v(y) = \inf_{Y\in{\cal Y}(y)}\E[V(Y_T)],\quad y>0.  
\end{equation}
Here $V$ is the convex conjugate function to $U$, that is
\begin{displaymath}
 V(y)=\sup_{x>0} \left\{U(x)-xy\right\}, \quad y>0,
\end{displaymath}
and ${\cal Y}(y)$ is the family of nonnegative supermartingales $Y$
that are dual to $\mathcal{X}(1)$ in the following sense
\begin{equation}
 \label{eq:Y(y)}
 {\cal Y}(y)=\{Y\geq 0:  Y_0=y \mbox { and } XY
\mbox { is a supermartingale for all } X\in\mathcal{X}(1)\}.
\end{equation}
Note that the set ${\cal Y}(1)$ contains the density processes of all
$\Q\in\mathcal{M}^e$ and  $Z\in \mathcal{Y}(1)$. 

The optimization problems \eqref{sup1} and \eqref{eq:dual} are
well studied. For example, it was shown in \cite {Kr-Scha 99} that under Assumption \ref{ass1} on the model and Assumptions \ref{ass2}, \ref{ass2'} on the utility function $U$, the  value functions $u$ and $v$ in \eqref{sup1} and \eqref{eq:dual}  are continuously differentiable on $(0,\infty)$ and they are conjugate
\begin{equation}
  \label{eq:conj_uv}
  v(y) = \sup_{x>0}\{u(x) - xy\}, \quad y>0.
\end{equation}
In addition, there exist unique optimizers
$\hat{X}(x,U)$ and $\hat{Y}(y,V)$ to \eqref{sup1} and \eqref{eq:dual} for all
$x>0$ and $y>0$. 
If $y=u'(x)$ then
\begin{equation}
\eqlabel{eq:dualXY}
  \hat{X}_T(x,U) =-V'( \hat{Y}_T(y,V) )
\end{equation}
and the product $\hat{X}(x,U)\hat{Y}(y,V)$ is a martingale, not only a supermartingale.  The value function in 
\eqref{eq:dual}
can also be  represented as as a supremum over the smaller set of densities of equivalent martingale measures
\begin{equation}\eqlabel{alternative}
v(y)=\inf_{\Q\in \mathcal{M}^e}\E[V(y\frac{d\Q}{d\IP})].
\end{equation} 
Before we prove our main results, we prove Proposition \ref{max-stoch-dom}:

\begin{proof}{~of Theorem \ref{max-stoch-dom}}:
(i) $\Longrightarrow$ (ii) If $f\in \mathcal{R}(S)$  then 
$f=p(f)+(H\cdot S )_T$ where $H\cdot S$ is a bounded martingale under $\Qmin$ and under each measure $\Q\in \cM^e$. Hence,
$$\E_{\Q}[f]=\E_{\Q^{(m)}}[f]=p(f) \text{~ for each~} \Q\in \mathcal{M}^e. $$
In particular, if 
$ L^{\infty}(\cG^N_T)  \subset \mathcal{R}(S)$ then  for each 
$f\in L^{\infty}(\cG^N_T) $ we have $\E_{\Q}[f]=\E_{Q^{(m)}}[f]$  or 
$\E[\frac{d\Q}{d\IP}f] =\E[\frac{d\Q^{(m)}}{d\IP}f] $. Taking  into account that $\frac{d\Q^{(m)}}{d\IP}$ is measurable with respect to 
$\cG^N_T$, we obtain
$$
\E[\frac{d\Q}{d\IP}|\cG^N_T]=\frac{d\Q^{(m)}}{d\IP}.$$

(ii) $\Longrightarrow$ (iii) is a well known property of second order stochastic domination, based on Jensen inequality. We include the short argument here for the sake of completeness.
 For any martingale measure $\Q\in \cM ^e$ we have 
\begin{equation}\eqlabel{eq:jensen}  \E[\psi(\frac{d\Q^{(m)}}{d\IP})] =
\E\left [\psi(\E[\frac{d\Q}{d\IP}|\cG^N_T])\right ] \leq 
\E\left[\E[\psi(\frac{d\Q}{d\IP})|\cG^N_T]\right] = E[\psi(\frac{d\Q}{d\IP})] ,
\end{equation}
for any convex function $\psi$ such that all the expectations above are well defined. This is equivalent to the second order stochastic domination relation 
\eqref{eq:stoch-dom}.

(iii) $\Longrightarrow$ (i): As just pointed out, the second order stochastic domination relation \eqref{eq:stoch-dom} is equivalent to inequality \eqref{eq:jensen} between the first and the last term, inequality that holds for all  convex "test" functions $\psi$.  If we consider the "test" function
 $\psi (\cdot)=V(y \cdot)$ for any $y>0$, where $V$ is the dual conjugate of an utility $U$
we obtain that 
 $$\E[V(y\frac{d\Q^{(m)}} {d\IP} ]\leq \E[V(y\frac{d\Q} {d\IP} )], \textrm{~for each~} \Q\in \mathcal{M}^e,$$ 
 which means that the infimum in \eqref{alternative} is attained by the martingale measure $\Qmin$. Comparing the optimization problems \eqref{eq:dual} and \eqref{alternative}, it is clear that \emph{once} the infimum in \eqref{alternative} is \emph{actually attained}, it also has to be the minimizer in the optimization problem \eqref{eq:dual}. Therefore, the minimizer 
 $\hat{Y}(y,V)$ is 
 (up to the multiplicative constant $y$) equal to 
 the density of the minimal martingale measure:
 \begin{equation}\eqlabel{unique:dual}
\hat{Y}_t(y,V)=y E[\frac{d\Q^{(m)}}{d\IP}|\mathcal{F}_t]=yZ_t, \ \ \ 0\leq t\leq T.
\end{equation}
Let  $h:\R\rightarrow [0,\infty)$ be a continuous  and strictly decreasing function such that
\begin{equation}\eqlabel{eq:der-dual-fct}
\lim_{y\rightarrow 0 }h(y)=\infty,\ \lim_{y\rightarrow \infty} h(y)=0\textrm{~and~} \int _0^{\infty} h(y)dy<\infty .
\end{equation}
If we denote by $V(y)=-\int _0^y h(u) du$, then the function $V$ is the conjugate function of a
\emph{bounded} convex utility function $U$, so that Assumption \ref{ass2'} is satisfied.
Taking $y=1$ in the optimization problem \eqref{eq:dual}, and taking into account \eqref{eq:dualXY}  together with \eqref{unique:dual} we conclude that the random variable
$h(\frac{d\Q^{(m)}}{d\IP})=-V'(\frac{d\Q^{(m)}}{d\IP})$ 
is the optimal terminal wealth for the investor having utility $U$ and initial wealth $x=-v'(1)$, i.e.
$h(\frac{d\Q^{(m)}}{d\IP})=X_T(-v'(1), U)$, therefore it is replicable by trading in $S$. Using an approximation argument, we obtain (i).
 \end{proof}
 \begin{remark}\eqlabel{rem:stoch-dom}
 From the proof of Proposition \ref{max-stoch-dom} it is easy to see  that items (i)-(iii) are also equivalent to relation \eqref{unique:dual}  and the (apparently stronger) stochastic domination condition
  $$\frac{d\Q^{(m)}} {d\IP} \succeq _2  Y_T \textrm{~for each~}Y\in \mathcal{Y}(1).$$ 
 \end{remark}
\begin{proof}{~of Theorem \ref{main1}}

Let us consider the optimizer $\hat{X}(x,U)$ in \eqref{sup1}. Under the hypotheses of Theorem \ref{main1} we know from Proposition \ref{max-stoch-dom} and Remark \ref{rem:stoch-dom} that relation \eqref{unique:dual} holds true. According to \eqref{eq:dualXY},
we have 
$$\hat{X}_T(x,U)=-V'(y\frac{d\Q^{(m)}}{d\IP}), \text{~for~}\ y=u'(x).$$
This implies  that $\hat{X}_T(x,U)$ is a positive,  $\mathcal{G}^N_T$-measurable and $\Q^{(m)}$-integrable random variable. The process $\hat{X}(x,U)$ is a true martingale under $\Q^{(m)}$.
Since $\hat{X}_T(x,U)$ is $\cG_T^N$-measurable and
$L^{\infty}(\cF^N_T)\subset \mathcal{R}(N,\cF)$, 
there exists a sequence of 1-dimensional integrands $k^n$ such that
$$\hat{X}_T(x,U)\wedge n=x_n+\int _0^T k^n_udN_u,$$
for each $n$, and $\int k^n dN$ is uniformly bounded in $t$ and $\omega$. Taking into account that the set of stochastic integrals which are true martingales under $\Qmin$ (identified with their last element) is closed in $L^1(\Omega, \mathcal{F}, \Qmin)$, 
we can let now $n\rightarrow \infty$ to conclude 
that there exists an $\cF$-predictable $k$ such that
$$\hat{X}(x,U)_t =x+\int _0^t k_udN_u,\ 0\leq t\leq T,$$
which finishes the proof. 
 \end{proof} 

\begin{proof}{~of Corollary \ref{cor:completeness}} The space $L^{\infty}(\mathcal{F}^N_T, \Omega, \IP)$ or, briefly, $L^{\infty}(\mathcal{F}^N_T)$
represents the set of all bounded path-dependent options on the num\'eraire portfolio $N$. Therefore, the assumption that the num\'eraire portfolio $N$ generates a complete market with respect to its own filtration $(\mathcal{F}^N_t)_{0\leq t\leq T}$ can be rewritten as
$L^{\infty}(\mathcal{F}^N_T)\subseteq \mathcal{R}(N, \mathcal{F}^N).$ Since, obviously,  
$\mathcal{R}(N, \mathcal{F}^N) \subset  \mathcal{R}(N)$ we obtain
$$L^{\infty}(\mathcal{F}^N_T)\subseteq \mathcal{R}(N).$$The above condition is  a stronger condition than the replicability condition (${\bf R}$), so, from Theorem \ref{main1} it follows that the (MFT) holds true for the model $S$ and the class of utilities $\mathcal{A}(S)$, and, in addition, the num\'eraire portfolio may serve as the mutual fund. We would like to point out that, under the hypotheses of Corollary
\ref{cor:completeness}, all utility maximizing agents will only invest in the num\'eraire porfolio and \emph{only} considering the information obtained by observing the evolution of the num\'eraire portfolio. In other words, the one-dimensional integrand in \eqref{eq MFT} is $(\mathcal{F}^N_t)_{0\leq t\leq T}$-predictable.
\end{proof}
\begin{remark}
The hypotheses of Corollary \ref{cor:completeness} may be replaced by assuming the the num\'eraire portfolio $N$ generates a complete market with respect to a filtration $(\mathcal{H}_t)_{0\leq t\leq T}$ such that
$$\mathcal{F}^N_t\subseteq \mathcal{H}_t\subseteq \mathcal{F}_t,\ \ \ 0\leq t\leq T.$$
As presented in the Proof of Proposition \ref{prp} below, the situation when $N$ generates a complete market with respect to a larger filtration $(\mathcal{H}_t)_{0\leq t\leq T}$  but \emph{not} with respect to its natural filtration $(\mathcal{F}^N_t)_{0\leq t\leq T}$ may, indeed, occur, so the remark extends the range of applicability of Corollary \ref{cor:completeness}.
\end{remark}

\begin{proof}{~of Theorem \ref{main2}}The  proof is very similar to the proof of (iii) $\Longrightarrow$ (i) in Proposition \ref{max-stoch-dom}.
Namelly, consider  the same kind of function $h:(0,\infty)\rightarrow \R$ as in \eqref{eq:der-dual-fct} and the associated dual function 
$V(y)=-\int_0^y h(z)dz$, as well as the convex conjugate utility function $U$. Since the weak completeness condition (${\bf WC}$) is satisfied,
according to  Theorem \ref{max-stoch-dom} and Remark \ref{rem:stoch-dom} we know that \eqref{unique:dual} holds true for any $y$, in particular when $y=1$. Using \eqref{eq:dualXY} we have that 
the random variable 
$h(\frac{1}{N_T}) =  h(\frac{d\Q^{(m)}}{d\IP})   = -V'(\frac{d\Q^{(m)}}{d\IP})$
is actually the terminal wealth of the investor with initial capital $-v'(1)$ and utility $U$, i.e. :
$$\hat{X}_T(-v'(1), U)=h(\frac{1}{N_T}).$$
 Since the (MFT) holds true for the model $S$ with respect to the set of all utilities $\mathcal{A}(S)$ and some mutual fund $M$ we conclude that  the random variable
$h(\frac{1}{N_T})$
can actually be  replicated by trading  in $M$ only, and the replication process (which is, in fact, $\hat{X}(-v'(1),U))$ is a martingale under $\Qmin$.  We can use now an approximation argument to  conclude that for any \emph{bounded and Borel measurable} function 
$g:\R \rightarrow \R$ we have that the random variable $g(N_T)$ is replicable by trading in $M$ only (and the stochastic integral is bounded) This means that
$$L^{\infty} (\cG^N_T)\subseteq \mathcal{R}(M),$$
so the proof is complete.
\end{proof}
\begin{remark}
In Theorem \ref{main1}, the num\'eraire portfolio $N$ can be replaced by  the 
wealth process $\hat{X}(x,U)$ of any rational investor (who possibly  has a utility function other than $U(x)=\ln x$). As a matter of fact,  both our main results amount to replicability  of random variables which are measurable with respect to the sigma algebra generated by the density of the greatest martingale measure in the sense of second order stochastic domination (in case such a measure exists). If such a measure exists, it obviously coincides with $\Qmin$.
\end{remark}
\begin{example}\eqlabel{ex:mf-needed} This example is a complement to Remark \ref{rem:mf-needed} and  shows that one can construct a complete market  where (MFT) holds true (so  condition (${\bf R}$) holds true, according to Theorem \ref{main2}) but the num\'eraire portfolio $N$  cannot be chosen as the mutual fund (or, in other words, condition (${\bf R^N}$) is not satisfied, according to Theorem \ref{main1}). The economic idea is the following: while the logarithmic investor is \emph{myopic}, always maximizing the expected logarithmic utility over the \emph{next} (infinitesimal) time period and therefore never investing in a martingale, some other investors may be willing to invest for some periods of time in martingales, in order to take advantage of the later benefits of being in a "better state" from the point of view of their expected utility at the final time-horizon.The construction goes as follows:
let $\ep_1, \ep_2$ be two independent $\{-1,1\}$-valued random variables with $\mathbf{P}[\ep_1=1]=\frac{1}{2}$ while $\mathbf{P}[\ep_2=1]=p\in \,]0,1[ \,\setminus \{\frac{1}{2}\}$. Define $S=(S_t)_{t=0}^2$ by
\begin{eqnarray*}
S_0 &=& 2 \mbox{ ~ (just to make } S \mbox{ positive)} ,\ \ \ \  \Delta S_1 = \ep_1\ \textrm{and}
\\ \Delta S_2 &=& 
\begin{cases}
\ep_2 & \mbox{ if }  \ep_1=1
\\ 0 & \mbox{ if } \ep_1=-1,
\end{cases}
\end{eqnarray*}
where $\Delta S_n=S_n-S_{n-1}$, $n=1,2$ are the increments of $S$. This defines a one-dimensional complete market $S$ for which (MFT) obviously  holds true. 

If we denote by $\widehat{H}^N=( \widehat{H}^N_1,\widehat{H}^N_2)$ the numbers of shares of the stock $S$ that the logarithmic maximizer holds between time $0$ and $1$, and time $1$ and $2$, respectively,  we can see (either by direct computation or a simple qualitative argument) that $\widehat{H}^N_1=0$  and 
$\widehat{H}^N_2=0$ if $\ep _1=-1$, $\widehat{H}^N_2\not= 0$ if $\ep _1=1$. On the other hand, for a generic utility function $U$ and endowment $x>0$,  the utility maximizing agent will take a non-trivial position over the first time period. 
In other words, using the same  notation 
 $\widehat{H}(x,U) =(\widehat{H}_1(x,U), \widehat{H}_2(x,U))$  for the  predictable process representing  the optimal number of shares, we may find a utility function $U$ (and an initial capital $x>0$) such that $\widehat{H}_1(x,U) \neq 0$. This obviously means that the num\'eraire portfolio $N$ cannot be chosen as mutual fund.

The fact that $\widehat{H}_1(x,U)\not= 0$ for some utility function $U$ can again be verified by direct computations (e.g. for the case $U(x) = \frac{x^\alpha}{\alpha}$) or by a more qualitative argument reproduced below.

Fix $x$ and $U$ and let $\widehat{X} (U,x) = x+ \widehat{H} \p S$ where $\widehat{H} = (\widehat{H}_1, \widehat{H}_2)$ is predictable.
Let us first look at the choice of $\widehat{H}_2$, which depends on $\ep_1$ and on the wealth $x+(\widehat{H}\p S)_1 = x+ \widehat{H}_0\Delta S_1$ of the investor at time $1$. Given this information the investor  solves the optimization  problem for the second period to determine $\widehat{H}_2$. 
The point is that we thus obtain a conditional indirect utility function $u_1 (x,\ep_1)$ which is defined as the optimal expected utility at time $T=2$, conditionally on the value of $\ep_1$ and wealth $x$ at time $t=1$. (For more details see, e.g., \cite{Scha 2002}.)
The optimization problem at time $t=0$, i.e., the determination of $\widehat{H}_1$, then may be viewed as a one period optimisation problem with horizon $t=1$ and with respect to the random utility function $u_1 (x, \ep_1)$.
The case of $U(x)=\log(x)$ is special, as in this case we have
\begin{equation}
u_1(x,\ep_1) = \log (x) + \mathrm{constant}(\ep_1),
\end{equation}
i.e., $u_1 (x,\ep_1)$ is a vertical shift of the logarithm, the constant depending on $\ep_1$ (but not on $x$). This is exactly the \emph{myopic} property of the logarithmic investor described in the beginning of the example, and can also be reformulated saying that the  marginal indirect utility $\frac{\partial}{\partial x} u_1(x,\ep_1)=x^{-1}$ does \emph{not} depend on $\ep_1$.

However, if we take a generic utility function $U$ (e.g., power or exponential), then $\frac{\partial}{\partial x} u_1 (x,\ep_1)$ will depend on $\ep_1$. Therefore, by a first moment argument, it will now be optimal to take a gamble on the random variable $\ep_1$, contrary to the case of the logarithm. 
More precisely, fix the endowment $x_0$ at time $t=0$. If $\frac{\partial}{\partial x} u_1 (x, \ep_1=1) |_{x=x_0} \neq \frac{\partial}{\partial x} u_1 (x, \ep_1=-1) |_{x=x_0}$ then the strategy $H_1=0$ cannot be optimal as one may exploit the difference of the marginal utilities.

We still remark that it is easy to reformulate Example \ref{ex:mf-needed}
in terms of continuous processes driven by a Brownian motion whose (infinitesimal) increments replace the independent random variables $\ep _1$ and $\ep_2$.
\end{example}
\begin{example}\label{example2}
In this example we will show that the "completeness" assumption (${\bf WC}$) is essential in Theorem \ref{main2}.
The basic idea of the (counter)example is that in  a one dimensional market, (MFT) always holds true
 in a trivial way,
 but the replicability condition $L^{\infty}(\cG^N_T)\subset \mathcal{R}(M)$ may fail, for any choice of the mutual fund $M$.

More precisely, 
consider a two dimensional Brownian motion $W=(W^{(1)}, W^{(2)}$ on the stochastic basis
$(\Omega, \mathcal{F}, (\mathcal{F}_t)_{0\leq t\leq T}, \IP)$ , where $(\mathcal{F}_t)_{0\leq t\leq T}$
is the filtration generated by the Brownian Motion $W$, and define the one-dimensional  stock price process by 
$$dS_t=S_t \{\arctg(W^{(2)}_t)^2dt+\arctg(W^{(2)}_t)dW^{(1)}_t\},\  S_0=1.$$
In this case, the num\'eraire portfolio equals the stock price process, i.e. $N=S$. Let us consider the random variable $f=\ln (N_T)$ and use Ito's formula to obtain 
 $$f=
 \int _0^T \frac{dS_t}{S_t}-\frac{1}{2}\int _0^T (\arctg(W^{(2)}_t)^2 dt.$$ 
Changing the measure $\IP$ to the minimal martingale measure $\Q ^{(m)}$ (defined by \eqref{xi-min} for $\gamma ^{(m)}_t=(\arctg (W^{(2)}_t,0), 0\leq t\leq T$) 
under which the second coordinate of the Brownian Motion, $W^{(2)}$, is left unchanged, we can easily see that the random variable 
$\int _0^T (\arctg(W^{(2)}_t)^2 dt$ cannot be replicated by trading in $S$, so, afortiori, the random variable $f$ cannot be replicated by trading in $S$.
Using now a simple density argument, we obtain that there exists  a positive integer $n$ such that 
$$\mathbb{I}_{\{-n\leq N_T\leq n\}}\ln (N_T) \notin \mathcal{R}(S),$$
so $f$ cannot be replicated by trading in any mutual fund $M$.
Since the random variable above is an element of  $L^{\infty } (\cG^N_T)$, the (counter)example  is complete.
\end{example}
\begin{proof}{~of Proposition \ref{prp}}
The construction is based on the well known idea that \emph{shrinking} the filtration may destroy the \emph{predictable representation property} (c.f. Problem 4.22 in \cite{Revuz-Yor}). 
As a matter of fact, the argument relies on a classical example of Tanaka, described below. 

Let $W=(W^{(1)},W^{(2)})$ be a 2-dimensional Brownian motion with its natural (right continuous, saturated) filtration $(\mathcal{F}_t)_{0\leq t\leq T} =(\mathcal{F}^W_t)_{0\leq t\leq T} $ on the probability space$(\Omega, \mathcal{F}_T,\IP)$.    
Define now the Levy transform of the first coordinate of the Brownian Motion
$$B^{(1)}_t=\int _0 ^t sgn (W^{(1)}_s) d W^{(1)}_s,$$
which is a one-dimensional Brownian Motion. At the risk of being somewhat pedantic, $B^{(1)}$ is a Brownian Motion with respect to the filtration $(\mathcal{F}_t)_{0\leq t\leq T} $, and, of course, also a Brownian Motion with respect to its own natural filtration $\cF^{B^{(1)}}=(\cF^{B^{(1)}})_{0\leq t\leq T}$ (which is right continuous and saturated).  The crucial point is the loss of information that occurs when passing from $W^{(1)}$ to $B^{(1)}$:  for each $0\leq t_0\leq T$, the random variable
$sgn (W^{(1)}_{t_0})$ and the process $(B^{(1)}_t)_{0\leq t\leq T}$ are independent under the probability measure $\IP$, and  the Brownian Motion $(B^{(1)}_t)_{0\leq t\leq T}$ is adapted to the filtration $(\cF ^{|W^{(1)}|})_{0\leq t\leq T}=(\cF^{B^{(1)}})_{0\leq t\leq T}$.

Fix $t_0\in (0, T)$ and consider the positive  process
$$h(t)=2+sgn( W^{(1)}_{t_0})1_{(t_0<t)},\ 0\leq t\leq T.$$
Note that the process $h$ is predictable with respect to the filtration $(\cF^{W^{(1)}}_t)_{0\leq t\leq T}$
(so, of course, predictable with respect to $(\cF_t)_{0\leq t\leq T}$).
On the triplet $(\Omega,(\mathcal{F}_t)  _{0\leq t\leq T},\IP )$ we consider the  two-dimensional stock price
$$\left \{ \begin{array}{ll}
dS^{(1)}_t=S^{(1)}_t [h^2(t)dt+h(t)dB^{(1)}_t]=S^{(1)}_t [h^2(t)dt+h(t) sgn (W^{(1)}_t)dW^{(1)}_t] , \  0\leq t\leq T\\
dS^{(2)}_t=S^{(2)}_t (\pi+\arctg (W^{(1)}_t))dW^{(2)}_t,\  \ 0\leq t\leq T .
\end{array} \right .$$ 
In other words, the first stock evolves like Geometric Brownian Motion driven by the the Brownian Motion $B^{(1)}$ and has volatility $2$ up to time $t_0$ and after that has either volatility $1$ or $3$, chosen independently of $B^{(1)}$, but adapted to $W^{(1)}$. We would like to point out  that, written in terms on $W$, all coefficients in the equation describing the evolution of the stock price are adapted to $W$. 
We also remark that $\cF^ {S^{(1)}}\subset \cF ^{W^{(1)}}$.

The unique equivalent martingale measure (for the two-dimensional stock price $S=(S^{(1)},S^{(2)})$ with respect to the filtration $(\cF_t)_{0\leq t\leq T}$) has  density 
$Z=1/S^{(1)}$, therefore, the portfolio num\'eraire process is the first stock, i.e.
$$N=S^{(1)}.$$
The second stock  $S^{( 2)}$ is not  traded by the logarithmic maximizer.  It serves to encode the information of both $W^{(1)}$ and $W^{(2)}$. As a matter of fact, 
$$\cF^W=\cF ^{S^{(2)}}\subset \cF ^S \subset \cF ^W,$$ so item (i) in the Proposition is satisfied.

Since $N=S^{(1)}$, we have $\cF^N \subset \cF ^{W^{(1)}}$. Using Girsanov Theorem (based on the predictable representation property of $W^{(1)}$ with respect to $\cF^{W^{(1)}}$), we can show that the only equivalent martingale measure for $N$ with respect to $\cF^{W^{(1)}}$ has density 
$Z=1/N=1/S^{(1)}$ (the same as the martingale measure for the two-dimensional stock $S$). This means that the reduced market $(N, \cF ^{W^{(1)}})$ is a complete market, which implies
$$  L^{\infty}(\mathcal{F}^N_T)   \subset \mathcal{R}(N,\cF ^{W^{(1)}})\subset \mathcal{R}(N, \cF),$$
so item (ii) is also satisfied.

For a fixed strike price $K>0$, let us consider $f=(K-N_T)^+$, the European put on the asset $N$ with maturity $T$. From the previous considerations ($(N, \cF ^{W^{(1)}})$ is a complete market) we conclude that there exist a price $p(f)>0$ and a $\cF^{W^{(1)}}$-predictable process $k$ such that 
the integral
$$I_t=p(f)+\int _0^t k_udN_u,\  \ 0\leq t\leq T,$$
is a bounded martingale under $\Q^{(m)}$  , and $f=I_T$. Assume that we can choose an integrand $k$ above which is $\cF^N$-predictable. From time $t_0$ to maturity $T$, the process $N=S^{(1)}$ evolves (except for the drift) like a Geometric Brownian Motion with volatility $D_{t_0}$, where 
we denote  by $D_{t_0}=2+sgn (W^1_{t_0})$. We emphasize that $D_{t_0}$ is  a $\cF^N_{t_0}$-measurable random variable, since $\cF^N$ is the \emph{saturated} filtration generated by $N=S^{(1)}$. If we denote be $BS(t,T, s, \sigma)$ the Black and Scholes price of the European put at time $t$, with maturity $T$, when the stock price is $s$ and volatility $\sigma$ (and, furthermore the interest rate is $r=0$ and call price is $K$), then the price of the above call $f$ at time $t_0$ is
$$I_{t_0}=BS(t_0,T, N_{t_0}, D_{t_0}).$$
The process $I$ is continuous and adapted to the filtration $\cF^N$, so $I_{t_0}$
is measurable with respect to the sigma-algebra
$\cF^N_{t_0 -}=\cF ^{B^{(1)}}_{t_0}$. However, this is absurd as $D_{t_0}$ and $\cF ^{B^{(1)}}_T$ are independent and the function $BS(t_0,T,s,\cdot)$ is strictly monotone. Therefore, we obtained by contradiction that the European option $f$ written on the asset $N$ cannot be replicated using integrands predictable with respect to $\cF^N$. In other words,  $f\in L^{\infty} (\cG^N_T)$ but $f \notin \mathcal{R}(N,\cF^N)$, so the proof is complete.
\end{proof}
\begin{remark}
Proposition \ref{prp} is sharp in two directions: the coefficients in \eqref{eqS} are adapted to the driving Brownian Motion, i.e. $\cF= \cF ^W$,
and the filtration is not larger than the one generated by the stock itself, which means $\cF =\cF ^S$. If we do not insist in  constructing a sharp example in the sense that $\cF^S=\cF^W$ as in item (i), then a one dimensional market following the dynamics of $S^1$ is sufficient for our purposes.
\end{remark}
\begin{proof}{ of Theorem \ref{propX2}}  
Fix two utility functions $U_1, U_2\in \mathcal{U}$. 
The proof is based on the construction of a 
(double indexed)  sequence of financial markets. The price process will be fixed, along with the risk-neutral measure, the only variable quantity being the objective probability measure.

Consider a two dimensional Brownian motion $W=(W^1, W^2)$ on the stochastic basis
$(\Omega, \mathcal{F}, (\mathcal{F}_t)_{0\leq t\leq T}, \Q)$ , where $(\mathcal{F}_t)_{0\leq t\leq T}$
is the saturated filtration generated by the Brownian Motion. We define the  two-dimensional price process by $S=(S^1,S^2)=W$, i.e the stock price follows the two-dimensional Brownian Motion so that the model is complete, and $\Q$ is the unique equivalent martingale measure. The (sequence of) financial markets described above will be constructed by considering appropriate \emph{objective} measures $\IP\approx \Q$.

Consider the $(\Q, \mathcal{F}_t)$ square-integrable martingale 
$$L _t =\int _0^t \arctg (W^2_s)dW^1_s,\  \ 0\leq t\leq T,$$
and let $$\tau _n =\inf \{t, |L _t |=n\ \textrm{or}\ t=T\},\  \ n=1,2, \dots\  
\text{~and~} \alpha _n =L_{\tau _n},\  \  \alpha = L_T.$$
The main idea is to construct the objective measure such that, for fixed $x>0, n\in \mathbb{N}$ and  fixed $\varepsilon\in \mathbb{R}$ satisfying $|\varepsilon | \leq 
x/2n$  the optimal wealth process of the investor with utility $U_1$ starting at $x$ is
$$X(x)=x+\varepsilon L^{\tau _n}, \text{~i.e~} X_T(x)=x+\varepsilon \alpha _n >0.$$
This can be  done by choosing the objective measure $\IP^{x,n, \varepsilon}$ such that
$$\frac{d\Q}{d\IP^{x,n, \varepsilon}}=  c(n,\varepsilon )U_1'(x+\varepsilon \alpha _n),$$
where the normalizing constant $c(x,n,\varepsilon )$ is chosen such that 
$$\E _{\Q}[\frac{d\IP^{x,n, \varepsilon}}{d\Q}]=1$$
Note that there are no integrability problems since $U_1'(x+\varepsilon \alpha _n)$ is bounded above and bounded away from zero.
We denote by $\mathcal{H}^2(\Q)$ the set of $(\Q, \mathcal{F}_t)$ square-integrable martingales
and by $\mathcal{H}^2_n(\Q)$ the closed subspace  of elements $M\in \mathcal{H}^2(\Q)$  such that
$$M_t=M_0+\int _0^tH^1_sdW^1_s+\int _0^tH^2_sdW^2_s  \text{~and~} H^2_t  1_{\{t\leq \tau _n\}}=0, \  \  0\leq t \leq T .$$
In other words, up to an additive constant,  $\mathcal{H}^2_n(\Q)$ consists of those stochastic integrals where the  second component of the integrand is identical zero before time $\tau _n$.  As a limiting case, we denote by $\mathcal{H}^2_{\infty}(\Q)$ 
the closed subspace  of $M\in \mathcal{H}^2(\Q)$  such that
$$M_t=M_0+\int _0^tH^1_sdW^1_s ,  \  0\leq t \leq T .$$

We identify spaces of martingales with their terminal values in the usual way, so we may say that 
$\alpha =L$ and $\alpha _n=L^{\tau _n}$.

Fix $y>0$. Since the random variable $yU_1'(x+\varepsilon \alpha _n)$ is a multiple of the density of the unique martingale measure for the financial market $S, (\Omega, \mathcal{F}, (\mathcal{F}_t)_{0\leq t\leq T},\IP^{x,n, \varepsilon} )$ 
we conclude that 
$I_2(yU_1'(x+\varepsilon \alpha _n))$ is the terminal wealth of an optimal investment strategy for the investor with utility $U_2$ (here $I_2$ is the inverse of $U_2'$).  If we suppose that MFT applies to the model $S$ defined on  $(\Omega, \mathcal{F}, (\mathcal{F}_t)_{0\leq t\leq T},\IP^{x, n, \varepsilon} )$ with respect to the class 
 $\mathcal{U}$ which contains $U_1$ and $U_2$   (and $\Q$ is the unique martingale measure), we conclude that
$$I_2(yU_1'(x+\varepsilon \alpha _n)) \in \mathcal{H}^2_n(\Q)$$ 
(we compare the optimal strategy of the investor with utility $U_2$ which has terminal wealth 
$I_2(yU_1'(x+\varepsilon \alpha _n))$ with the optimal investment strategy of the investor with utility $U_1$ starting at $x$)
Letting $f(x, y)=I_2(yU_1'(x))$, we have obtained that, for fixed $y>0$,$n\in\mathbb{N}$, $x>0$ and $|\varepsilon | \leq 
x/2n$  we have
$$f(x+\varepsilon \alpha _n,y)\in \mathcal{H}^2_n(\Q).$$ 
For $x>0$ and $\varepsilon \leq x/2n$ we define the two-variable $\mathcal{H}^2(\Q)$-valued function
$$F(x,\varepsilon):= f(x+\varepsilon \alpha _n,y)\in \mathcal{H}^2_n(\Q)\subset  \mathcal{H}^2(\Q).$$
Using the bounded convergence theorem (since $\alpha _n$ is bounded and $f(\cdot, y)$ is $C^2$), we obtain that the function $F$ is two-times differentiable with respect to $\varepsilon$ as a $\mathcal{H}^2(\Q)$-valued function and 
$$\frac{\partial ^2}{\partial \varepsilon ^2}F(x,\varepsilon) |_{\varepsilon =0}=
\frac{\partial ^2 f}{\partial x^2}(x,y) (\alpha _n)^2 \in \mathcal{H}^2 (\Q).$$ 
Since $\mathcal{H}^2_n(\Q)$ is a closed subspace of $\mathcal{H}^2(\Q)$, we conclude that
$$
\frac{\partial ^2 f}{\partial x^2}(x,y) (\alpha _n)^2 \in \mathcal{H}^2_n (\Q).$$ Taking into account that 
$\frac{\partial ^2 f}{\partial x^2}(x,y)$ is a scalar (real number) we can let $n\rightarrow \infty$ to obtain 
\begin{equation}\label{eq:f"}
\frac{\partial ^2 f}{\partial x^2}(x,y) \alpha ^2 \in \mathcal{H}^2_{\infty}(\Q).
\end{equation}
However, according to Ito formula
$$\alpha ^2=(L_T)^2=2\int _0^TL_t \arctg (W^2_t)dW^1_t+ \int _0^T (\arctg (W^2_t))^2 dt,$$ 
so 
$$\alpha ^2 \notin  \mathcal{H}^2_{\infty}(\Q).$$
The above relation, together with \eqref{eq:f"}, shows that 
 $$ \frac{\partial ^2 I_2(yU_1'(x))}{\partial x^2}= \frac{\partial ^2 f}{\partial  x^2}(x,y) =0,\  \  \text{~for all~} x>0, y>0.$$ 
This means that the function $I_2(yU_1'(\cdot))$ is linear, or $I_2(yU_1'(x))=a(y)x+b(y)$ for any $x,y>0$. Since $I_2, U_1':(0,\infty)\rightarrow (0,\infty)$ are strictly decreasing bijective functions, we conclude that $x\rightarrow I_2(yU_1'(x))$ is a strictly increasing and bijective function from $(0,\infty)$ to 
$(0, \infty)$, so $b(y)=0$ and $a(y)>0$ for each $y>0$. Considering $x=I_1(z)$, we obtain
$I_2(yz)=a(y)I_1(z), \textrm{~for each~}x,z>0$, which can be rewritten  as
$$\frac{I_2(yz)}{I_1(z)}=a(y),\textrm{~for each~}x,z>0.$$
Differentiating with respect to $z$, we obtain
$$\frac{yI_2'(yz)I_1(z)-I_2(yz)I_1'(z)}{(I_1(z))^2}=0,$$
or
$$\frac{I_2'(yz)}{I_2(yz)}=\frac{1}{y}\frac{I_1'(z)}{I_1(z)}\textrm{~for each~}x,z>0.$$
Now, we first consider $z=1$ and allow $y$ to vary over $(0,\infty)$ and then  we consider $y=1$ and allow $z$ to vary over $(0,\infty)$ to obtain 
$$\frac{I_2'(y)}{I_2(y)}=\frac{I_1'(1)}{I_1(1)}\frac{1}{y}=\frac{I_1'(y)}{I_1(y)}\textrm{~for each~}y>0.$$
Taking into account that $\frac{I_1'(1)}{I_1(1)}<0$ is a constant, the conclusion of the Theorem follows by integration.
\end{proof}

\bibliographystyle{amsplain}

\end{document}